\documentclass[fleqn,12pt,twoside,espcrc1]{article}
\usepackage[headings]{espcrc1}
\usepackage{wrapfig}
\readRCS
$Id: espcrc1.tex,v 1.2 2004/02/24 11:22:11 spepping Exp $
\usepackage{graphicx}
\usepackage[figuresright]{rotating}

\newcommand{\AmS}{{\protect\the\textfont2
  A\kern-.1667em\lower.5ex\hbox{M}\kern-.125emS}}

\title{Identical Particle Correlations in STAR}

\author{Z. Chaj\c{e}cki\address[MCSD]{The Ohio State University, 191 W. Woodruff Avenue, Columbus, Ohio 43210, USA}
        (for the STAR\thanks{For the full list of STAR authors and acknowledgments, 
	see appendix 'Collaborations' of this volume.} Collaboration)}
       
\runtitle{Identical Particle Correlations in STAR}
\runauthor{Z. Chaj\c{e}cki (for the STAR Collaboration)}

\begin{document}

\maketitle

\begin{abstract}

Preliminary results of identical-particle correlations
probing the geometric substructure of the particle-emitting source at RHIC are presented.
An $m_T$-independent scaling of pion HBT radii from large (central Au+Au) to small (p+p) 
collision systems naively suggests comparable flow strength in all of them.
Multidimensional correlation functions are studied in detail using a spherical
decomposition method. 
In the light systems, the presence of significant long-range non-femtoscopic
correlations complicates the extraction of HBT radii.

\end{abstract}

\section{Introduction}
Particle interferometry is a useful technique that provides information on the space-time 
properties of nuclear matter created in high energy collisions (for the latest review articles
see \cite{Lisa2005,Review}). 
In this paper results of particle correlations in p+p, d+Au and Cu+Cu collisions at 
$\sqrt{s_{NN}}=200$ GeV and Au+Au at $\sqrt{s_{NN}}=200$ and $62$ GeV registered by the STAR 
experiment are presented. Rich data statistics and the wide acceptance of the detectors
gives an opportunity to do a multidimensional femtoscopic analysis.
The main focus of this article is on the transverse mass ($m_T=\sqrt{k_T^2+m_{\pi}^2}$) dependence of the HBT 
radii for different system sizes. This allows the study  of the dynamics of the source and the conditions of the collision.
Importantly, direct comparison between the $m_T$ dependence of radii from p+p and A+A 
collisions is possible for the first time.

\begin{figure}[htb]
\begin{minipage}[t]{85mm}
  \includegraphics[width=85mm]{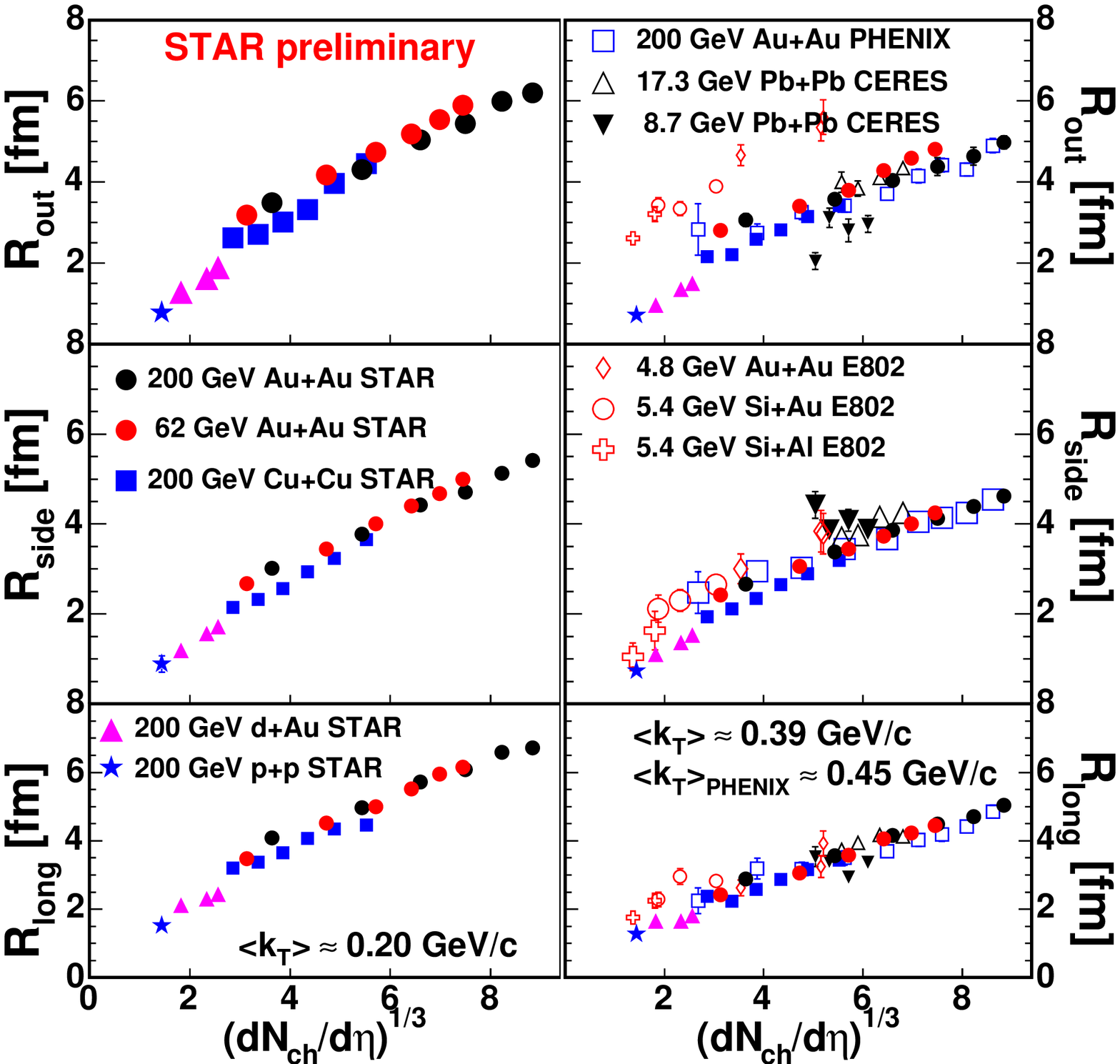}
\vspace*{-1.3cm}
  \caption{Femtoscopic radii dependence on the number of charged particle.}
  \label{fig:UniversalScaling}
\end{minipage}
\hspace{\fill}
\begin{minipage}[t]{75mm}
\includegraphics[width=75mm]{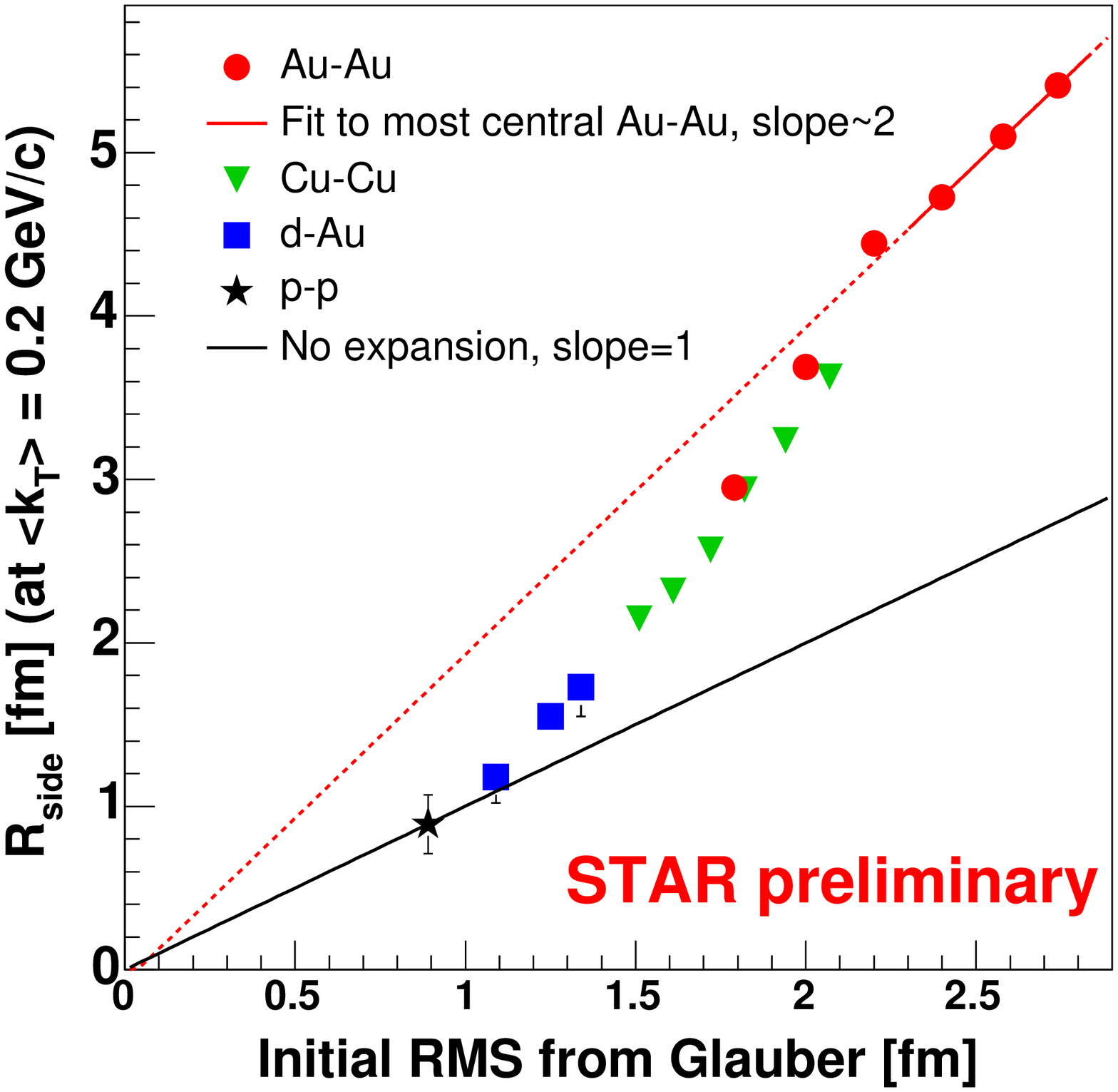}
\vspace*{-1.3cm}
\caption{Final size of the source vs. initial radii calculated from Glauber model. STAR Au+Au data from \cite{Mercedes2004}.}
  \label{fig:SystemExpansion}
\end{minipage}
\vspace*{-0.8cm}
\end{figure}

\section{Multiplicity Scaling and Expansion}

Figure \ref{fig:UniversalScaling} presents the HBT radii 
dependence on $(dN_{ch}/d\eta)^{1/3}$ ($dN_{ch}$ - number of charged particles) for different colliding systems 
at different energies of the collisions. The motivation for studying such  
a relation is its connection to the final state geometry 
through the particle density at freeze-out.  
All STAR results, from p+p, d+Au, Cu+Cu and  Au+Au collisions, are combined on the left panel
of this figure and, as seen, all radii exhibit a scaling with $(dN_{ch}/d\eta)^{1/3}$.
On the right panel STAR radii, this time for different
range of $k_T$,
 are plotted together 
with AGS/SPS/RHIC systematics \cite{Lisa2005}. 
It is impressive that the radius parameters $R_{side}$ and 
$R_{long}$ follow the same curve for different collisions  over a wide range 
of energies and, as it was checked, this observation 
is valid for all $k_T$ bins studied by STAR.
It is a  clear signature 
that the multiplicity is a scaling variable that drives these HBT  radius parameters. 
$R_{out}$ mixes space and time information. Therefore it is unclear
whether to expect a simple scaling with the final state geometry.

Despite the approximately linear dependence of $R_{side}$ and $R_{long}$
on $d(dN_{ch}/d\eta)^{1/3}$, because of the finite intercepts of the linear
relationships \cite{Lisa2005,PhenixPRL2}, the data do not imply freeze-out
at a constant density.  The fact that the radii scale at all (linearly or not)
indicates that the sideward and longitudinal radius is determined only
by multiplicity, independent of energy, colliding system, or impact
parameter \cite{Stock,CsorgoCsernai,Lisa2005}.  This scaling, however, breaks down
at much lower energies, when baryons constitute a significant fraction
of the freeze-out system \cite{Stock,Ceres,Lisa2005}.



Figure \ref{fig:SystemExpansion} suggests that the relationship between initial and final
geometry is not trivial, however.  There, the final RMS of the source,
estimated by $R_{side}$ at low $k_T$ \cite{BW}, is plotted versus the initial RMS
of the overlap region, estimated with a Glauber calculation.  The system
generated in central Au+Au collisions undergoes a two-fold expansion,
while those from the most peripheral d+Au and p+p collisions expand
little.  Does this imply that small systems are less explosive than
large ones?  This is explored in the next Section.

\section{Transverse Mass Dependence of HBT Radii : p+p vs Au+Au}

\begin{figure}[htb]
\begin{minipage}[t]{75mm}
  \includegraphics[width=75mm]{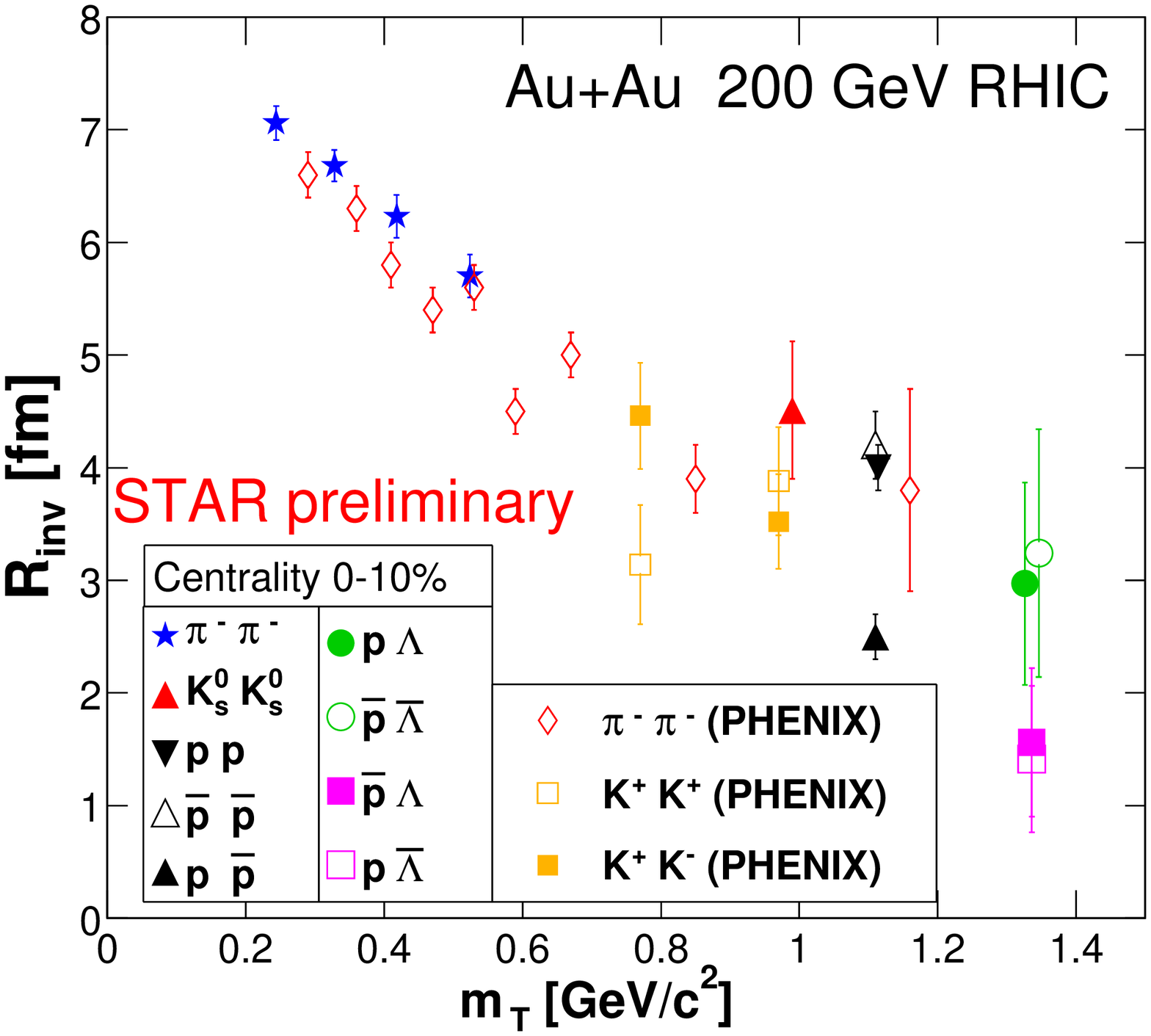}
\vspace*{-1.3cm}
  \caption{$m_T$ dependence of $R_{inv}$ for different particles. PHENIX data from \cite{PhenixQM04}.}
\label{fig:Rinv}
\end{minipage}
\hspace{\fill}
\begin{minipage}[t]{80mm}
\includegraphics[width=80mm]{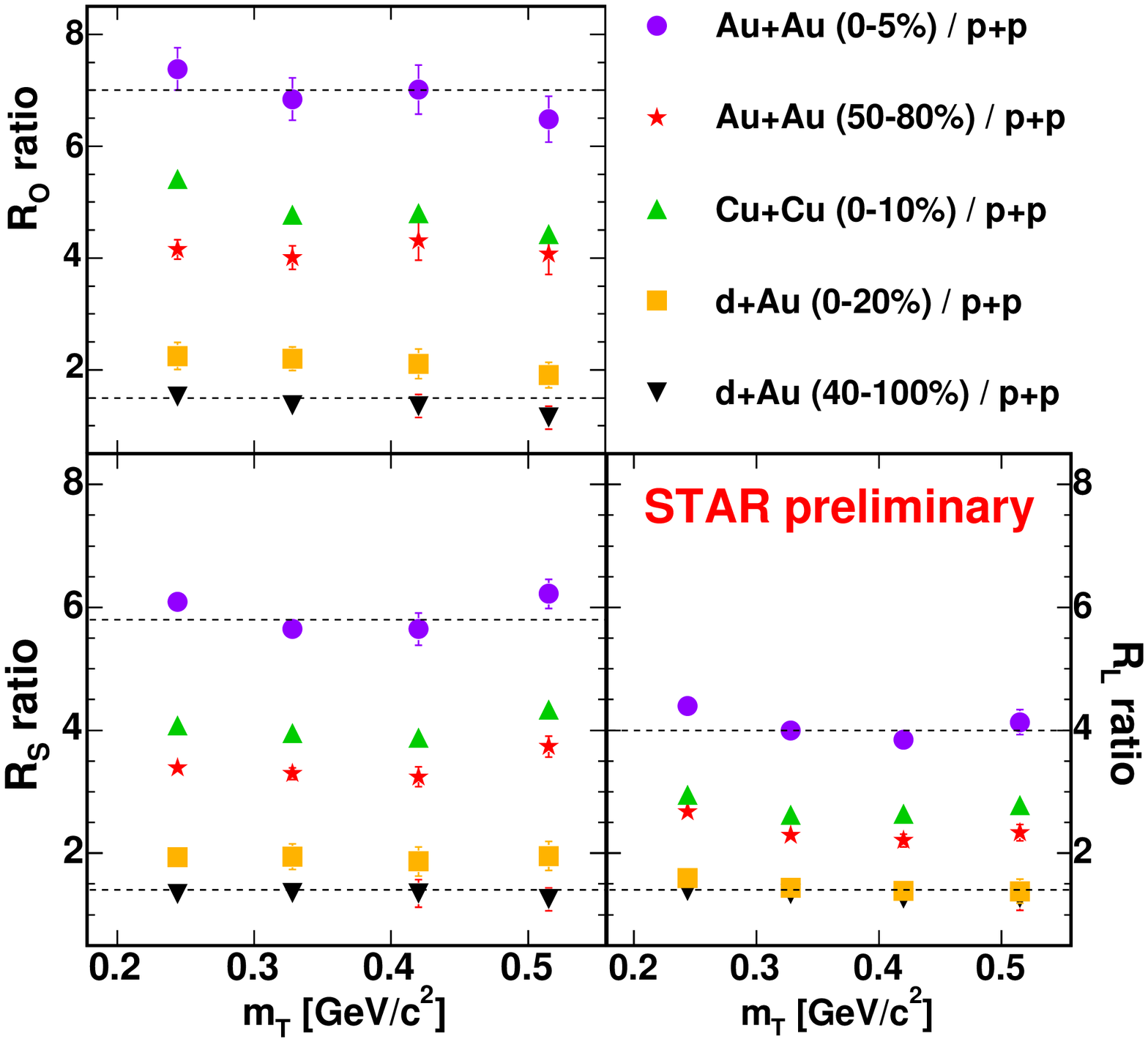}
\vspace*{-1.3cm}
\caption{Ratio of HBT radii from Au+Au, Cu+Cu and d+Au by p+p collisions at $\sqrt{s_{NN}}=200$ GeV.}
\label{fig:kTscaling}
\end{minipage}
\vspace*{-0.8cm}
\end{figure}

  The $m_T$ dependence of the femtoscopic radii in heavy ion collisions is
usually attributed to collective flow of a bulk system, and recent STAR
results for central Au+Au collisions support this picture \cite{Mercedes2004}.  In a flow
scenario, an approximately ``universal'' $m_T$ dependence should apply to all
particle types.  This is in fact observed in Figure \ref{fig:Rinv}, in which 
one-dimensional radii from pion \cite{PhenixQM04}, charged kaon \cite{PhenixQM04},
 neutral kaon \cite{BekeleSQM}, 
proton and anti-proton \cite{GosQM05}, and proton-$\Lambda$ \cite{ChaloupkaQM05} correlations are plotted.
Correlations between particles with very different masses also show
characteristic signals of collective flow \cite{STARkpi}.

  Decreasing $m_T$ dependences of the radii have previously been reported
for hadron \cite{Agababyan} and $e^++e^-$ \cite{Alexander} collisions.  Until the STAR analysis, a
direct comparison of the $m_T$ systematic from hadron collisions with those
from A+A collisions had been hampered by the use of quite different
parameterizations in the different experiments.
  As shown in Figure \ref{fig:kTscaling}, the {\it ratio} of HBT radii from Au+Au
collisions with those from p+p collisions are approximately independent
of $m_T$.  It might be that proposed explanations for the $m_T$ dependence in
small systems, including Heisenberg-type relations \cite{Alexander} and inside-outside
cascade scenarios \cite{Bialas}, replicate the $m_T$ dependence from bulk collective
flow.  This would be a surprising and unfortunate coincidence, suggesting
that dynamical signals in the space-time sector cannot distinguish very
different underlying physics.  Alternatively, as proposed by Cs{\"o}rg{\H o} {\it et al.} \cite{Csorgo2004}, 
it could be that p+p collisions generate a thermalized {\it bulk system}, 
similar to that created in heavy ion collisions.
  Further theoretical work on this important question would be most
welcome.  However, as it is mentioned in the next Section, further work is
also required on the experimental side.

\section{The effect of non-femtoscopic correlations in low-multiplicity events.}

HBT radii may arise from Gaussian fits to
the three-dimensional correlation function \cite{Lisa2005}.  But for low-multiplicity
systems, unaccounted-for structures in the correlation function itself are observed.
  In particular, the value of the correlation function $C(\vec{q})$
for large $|\vec{q}|$ (larger than the scale of quantum statistics or Coulomb
interactions) does not approach a common $\vec{q}$-independent value.
  Thus,
non-femtoscopic correlations (e.g. due to momentum conservation) are becoming
significant here; they have been also observed earlier, in elementary particle collisions, see e.g. \cite{Agababyan}.

\begin{wrapfigure}{rt}{70mm}
\vspace*{-1.8cm}
 \begin{center}
   \includegraphics[width=70mm]{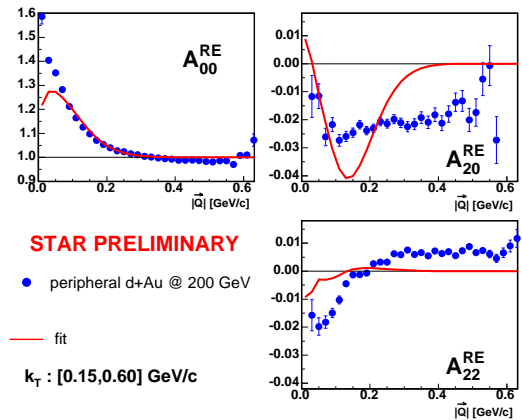}
 \end{center}
\vspace*{-.9cm}
 \caption{First three non-vanished components of the spherical harmonic decomposition 
   of the correlation function for peripheral d+Au collisions.}
 \label{fig:SH}
\vspace*{-1.4cm}
\end{wrapfigure}

  The spherical decomposition of the correlation function in $\vec{q}$-space
has been proposed \cite{Chajecki2005} as a sensitive measure of long-range non-femtoscopic
correlations.  The lowest-$l$ components $A_{l,m}$ of this procedure are
shown in Figure \ref{fig:SH}.  For all correlations of femtoscopic origin (Gaussian or
not), $A_{l\neq 0,m}$ must vanish \cite{Chajecki2005} for large $\vec{q}$; clearly other correlations are present
in STAR data.

  While various possible sources of these correlations are still being explored,
it is interesting to simply account for the long-range correlations by adding
ad-hoc terms to the Gaussian fits, which produce constant values of $A_{2,0}$
and $A_{2,2}$.  It was found that the ratios shown in Figure 4 remain
$m_T$-independent, though their values shift somewhat.

\section{Conclusions}

The results of pion interferometry for all energies and colliding systems at RHIC
 have been presented. 
In agreement with data at SPS and AGS, STAR 
indicates that the multiplicity is the scaling variable that determines 
the size of the source at freeze-out.
Perhaps surprisingly, the $m_T$ dependence of HBT radii appears to be independent of 
collision species or multiplicity.
Finally, a problem with the baseline of the correlation function for low multiplicity 
collisions has been reported, and a promising tool based on the spherical 
harmonic decomposition of the correlation function has been used in order to address it.
The physics of this structure remains under investigation.

\end{document}